\documentclass[prd,aps,showpacs,nofootinbib,preprint,eqsecnum]{revtex4}
\usepackage{graphicx,color,amsmath,amsxtra}
\usepackage{amssymb}
\usepackage[english]{babel}
\usepackage{amsfonts}
\allowdisplaybreaks[4]

\begin{document}

\title{Phantom crossing dark energy in Horndeski's theory}

\author{Jiro Matsumoto\footnote{Email: jmatsumoto@ntu.edu.tw}}
\affiliation{Leung Center for Cosmology and Particle Astrophysics, National Taiwan University, Taipei 10617, Taiwan\\
Institute of Physics, Kazan Federal University, Kremlevskaya Street 16a,
Kazan 420008, Russia}
	
\begin{abstract}
The $\Lambda$CDM model is a remarkably successful model which is consistent with the observations of cosmic microwave background radiation (CMB), baryon acoustic oscillation (BAO), and large scale structure of the Universe. 
However, the discrepancy in the value of $H_0$ between the local observations and PLANCK observation of CMB was recently pointed out. 
One of the way to ease the discrepancy is to introduce phantom dark energy instead of the cosmological constant. 
While, phantom dark energy often suffer from the instabilities. We will investigate the general solution to overcome the difficulty of phantom dark energy and construct some particular models which have 
a phantom crossing and can be consistent with the observations. 
\end{abstract}
	

\maketitle
\section{introduction \label{sec1}}
The accelerated expansion of the Universe was discovered by the observations of type Ia supernovae in 
late 1990s \cite{Riess:1998cb,Perlmutter:1998np}, and it is now also supported by the other observations:
Cosmic Microwave Background Radiation (CMB) \cite{Komatsu:2010fb,Ade:2013zuv,Ade:2015xua},
Baryon Acoustic Oscillations (BAO)
\cite{Percival:2009xn,Blake:2011en,Beutler:2011hx,Cuesta:2015mqa,Delubac:2014aqe},
and so on. 
To realize the accelerated expansion in the homogeneous and isotropic Universe, it is necessary to introduce some energy components with the equation of 
state parameter $w=p / \rho$ which is less than $-1/3$ into the Einstein equations. 
Dark energy is a hypothetical energy which has such a property. The most famous 
model of dark energy is the cosmological constant $\Lambda$, and the $\Lambda$ Cold Dark Matter ($\Lambda$CDM) model is known as 
the standard model in cosmology. The $\Lambda$CDM model is a simple model but almost consistent with 
all of the astronomical observations. 
However, the recent observations of supernovae and Cepheid variables \cite{Riess:2016jrr,Riess:2011yx} and 
the observation of CMB by PLANCK satellite \cite{Ade:2013zuv,Ade:2015xua} show there is over $2 \sigma$ discrepancy in the obtained values of $H_0$. 
This result would imply that dark energy is not constant but dynamical \cite{Marra:2013rba,DiValentino:2016hlg}. 
Moreover, it is also shown that dynamical dark energy which have a phantom crossing is favored if we take the other observations into account \cite{Zhao:2017cud}. 
Here, phantom crossing is a phenomenon that equation of state parameter $w$ dynamically crosses over the value $-1$. 

Whereas, a famous model of dynamical dark energy, quintessence model \cite{Peebles:1987ek,Ratra:1987rm,Chiba:1997ej,Zlatev:1998tr}, can not realize phantom crossing without instability, because phantom crossing is only realized when the sign of kinetic term flips \cite{Vikman:2004dc}.  
Horndeski's theory \cite{Horndeski} is known as a general theory of scalar-tensor theory including quintessence model as a special case. 
In Horndeski's theory, it is also known that phantom dark energy is, in some cases, realized without instability (e.g. see \cite{DeFelice:2010pv,Deffayet:2010qz}). 
We will investigate the general conditions for realizing stable phantom dark energy in Horndeski's theory in this paper. 
The contents of the paper are as follows;
general background equations and sound speeds in Horndeski's theory are given in Sec.~\ref{sec2}, 
the general conditions for stable phantom dark energy are derived in Sec.~\ref{sec3}, 
some examples of phantom crossing dark energy are given in Sec.~\ref{sec4},
concluding remarks are in Sec.~\ref{sec5}.
We use Natural units, $\hbar =c = k_\mathrm{B}=1$, 
and gravitational constant $8 \pi G$ is denoted by
${\kappa}^2 \equiv 8\pi/{M_{\mathrm{Pl}}}^2$ 
with the Planck mass of $M_{\mathrm{Pl}} = G^{-1/2} = 1.2 \times 10^{19}$GeV in the following.

\section{Horndeski's theory \label{sec2}}
The action in Horndeski's theory is given by \cite{Horndeski,Deffayet:2011gz,Kobayashi:2011nu}
\begin{equation}
S_H=\sum ^5 _{i=2} \int d ^4 x \sqrt{-g} {\cal L}_i,
\end{equation}
where
\begin{align}
{\cal L}_2 &= K(\phi , X), \\
{\cal L}_3 &= -G_3(\phi , X) \Box \phi , \\
{\cal L}_4 &= G_4(\phi , X)R + G_{4 X} \left [ ( \Box \phi )^2 - (\nabla _\mu \nabla _\nu \phi)^2 \right ], \\
{\cal L}_5 &= G_5 (\phi , X)G_{\mu \nu} \nabla ^{\mu}\nabla ^{\nu} \phi - 
\frac{G_{5X}}{6} \left [ (\Box \phi)^3 - 3 (\Box \phi) (\nabla _ \mu \nabla _ \nu \phi)^2 +2 (\nabla _ \mu \nabla _ \nu \phi)^3 \right ]. 
\end{align}
Here, $K$, $G_3$, $G_4$, and $G_5$ are generic functions of 
$\phi$ and $X=- \partial _\mu \phi \partial ^\mu \phi /2$, and the subscript $X$ means derivative with respect to $X$. 
The total action we will consider is sum of $S_H$ and the action of matter fluid $S_\mathrm{matter}$, 
which contain baryons and cold dark matter. 
The background equations of the Universe are given by 
assuming homogeneity and isotropy of the metric. 
Substituting $\phi = \phi (t)$ and the metric $ds^2 = -N^2 (t) dt^2 + a^2 (t)dx^2$ into the action, 
subsequently, variating the action with respect to $N(t)$ gives \cite{Kobayashi:2011nu}
\begin{equation}
\rho _\mathrm{matter} + \sum ^5 _{i=2} {\cal E}_i=0 \label{FL1}, 
\end{equation}
where
\begin{align}
{\cal E}_2 &= 2XK_X - K, \\
{\cal E}_3 &=6X \dot \phi HG_{3X} -2X G_{3 \phi}, \\
{\cal E}_4 &= -6H^2 G_4 +24 H^2 X (G_{4X}+XG_{4XX})-12HX \dot \phi G_{4 \phi X}
-6H \dot \phi G_{4 \phi}, \\
{\cal E}_5 &= 2H^3 X \dot \phi (5 G_{5X}+ 2XG_{5XX}) -6 H^2 X (3G_{5 \phi} +2XG_{5 \phi X}), 
\end{align}
and $\rho _\mathrm{matter}$ is the energy density of matter. Here, 
$H=\dot a /a$ is the Hubble rate function and the dot means derivative with respect to time. 
While, variation with respect to $a(t)$ yields 
\begin{equation}
p _\mathrm{matter} + \sum ^5 _{i=2} {\cal P}_i=0 \label{FL2}, 
\end{equation}
where
\begin{align}
{\cal P}_2 &= K, \\
{\cal P}_3 &= -2X \Big (  G_{3 \phi}+ \ddot \phi G_{3X} \Big), \\
{\cal P}_4 &= 2(3H^2 +2 \dot H) G_4 -4 H^2 X \bigg ( 3+ \frac{\dot X}{HX} +2 \frac{\dot H}{H^2} \bigg ) G_{4X} \nonumber \\
&-8HX \dot X G_{4XX} +2(\ddot \phi +2H \dot \phi ) G_{4 \phi }+4XG_{4 \phi \phi} + 4X(\ddot \phi -2H \dot \phi ) G_{4 \phi X}, \\
{\cal P}_5 &= -2X(2H^3 \dot \phi +2H \dot H \dot \phi +3H^2 \ddot \phi) G_{5 X}-4 H^2 X^2 \ddot \phi G_{5XX} \nonumber \\
&+4HX(\dot X -HX)G_{5 \phi X} +2 H^2 X \bigg ( 3+ 2\frac{\dot X}{HX} + 2 \frac{\dot H}{H^2} \bigg )G_{5 \phi} 
+4HX \dot \phi G_{5 \phi \phi}, 
\end{align}
and $p _\mathrm{matter}$ is the pressure of matter.
The above two Eqs.~(\ref{FL1}) and (\ref{FL2}) correspond to the Friedmann equations.  
The equation of motion of the scalar field is given by variating the action with respect to $\phi (t)$: 
\begin{equation}
\frac{1}{a^3} \frac{d}{dt} (a^3 J) = P_\phi, \label{FE}
\end{equation}
where
\begin{align}
J=& \dot \phi K_X + 6HXG_{3X} -2 \dot \phi G_{3 \phi} + 6H^2 \dot \phi (G_{4X}+2XG_{4XX})-12HXG_{4 \phi X} \nonumber \\
&+ 2H^3 X(3G_{5X} + 2XG_{5XX}) -6H^2 \dot \phi (G_{5 \phi} + XG_{5 \phi X}),
\end{align}
\begin{align}
P_{\phi} =& K_\phi -2X (G_{3 \phi \phi} + \ddot \phi G_{3 \phi X}) + 6(2H^2 + \dot H) G_{4 \phi} +6H(\dot X +2HX) G_{4 \phi X} \nonumber \\
&- 6H^2 X G_{5 \phi \phi} + 2H^3 X \dot \phi G_{5 \phi X}.
\end{align}
Equations (\ref{FL1}), (\ref{FL2}), and (\ref{FE}) control background evolution of the Universe. 
In the same manner as quintessence model, Eqs.~(\ref{FL2}) and (\ref{FE}) are equivalent when Eq.~(\ref{FL1}) holds. 
Equations (\ref{FL1}) and (\ref{FL2}) can be rewritten as well-known form: 
\begin{align}
3H^2 = \kappa ^2 (\rho _\mathrm{matter} + \rho _\phi), \\ 
-3H^2 -2 \dot H = \kappa ^2 (p_\mathrm{matter} + p_\phi),
\end{align}
if we define $\rho _{\phi}$ and $p _\phi$ as 
\begin{equation}
\rho _{\phi} \equiv \sum ^5 _{i=2} {\cal E}_i + \frac{3H^2}{\kappa ^2}, \qquad 
p_{\phi} \equiv \sum ^5 _{i=2} {\cal P}_i - \frac{1}{\kappa ^2} (3H^2 + 2 \dot H). \label{rhop}
\end{equation}
We will use Eq.~(\ref{rhop}) as the definitions of effective energy density and effective pressure. 

The perturbative behavior of the Universe can be described if we employ metric perturbations, the perturbation of $\phi$, 
and that of energy-momentum tensor of matter. 
Sound speed is an important quantity for understanding the dynamics of perturbation quantities, 
because the propagating speed of perturbation quantities are determined by sound speed if it is not zero. 
The sound speed for tensor perturbations is expresses as \cite{Kobayashi:2011nu}
\begin{equation}
c_T ^2 = \frac{G_4 -XG_{5 \phi}-XG_{5X} \ddot \phi}{G_4 -2XG_{4X}-X(G_{5X}\dot \phi H - G_{5 \phi})}. 
\label{ct}
\end{equation}
Equation (\ref{ct}) shows that the sound speed for tensor perturbations is independent from 
the functions $K(\phi ,X)$, $G_3 (\phi , X)$, and matter components. 
If the terms $XG_{5 \phi}$, $XG_{4X}$, $\cdots$  are relevant for the evolution of the Universe, then 
they should be same order as $G_4$ as seen from Eqs.~(\ref{FL1}) and (\ref{FL2}). 
In this case, $c_T^2$ is deviate from $1$ except for some special cases. 
While, the recent observation of gravitational wave GW170817 \cite{TheLIGOScientific:2017qsa}
and its electromagnegic counter parts \cite{Monitor:2017mdv,GBM:2017lvd,Coulter:2017wya} showed that the speed of 
gravitational wave should satisfy 
\begin{equation}
\vert c_T^2 -1 \vert \lesssim 10^{-15} 
\end{equation}
in relatively recent Universe. 
This bound means that the speed of gravitational wave should be almost same as that of electromagnetic wave not only 
around stellar objects but also in void region.  
Therefore, it is natural to think that the terms proportional to $G_{4X}$, $G_{5 \phi}$, and $G_{5 X}$ are 
not relevant for the current accelerated expansion of the Universe.  
In the following, we treat $G_4(\phi , X)$ and $G_5(\phi, X)$ as $G_4(\phi)$ and $G_5(\phi , X)=0$. Here,  
$G_5 (\phi , X)$ is not expressed as constant but as $0$ because constant $G_5$ does not contribute to Eqs.~(\ref{FL1}), (\ref{FL2}), and (\ref{FE}).  
Further discussions for the constraints from gravitational wave detection GW170817 for Horndeski's theory are given in Refs.
~\cite{Creminelli:2017sry,Sakstein:2017xjx,Ezquiaga:2017ekz,Baker:2017hug,Arai:2017hxj,Langlois:2017dyl,Lombriser:2015sxa,Lombriser:2016yzn}. 

The sound speed for scalar mode is written as 
\begin{align}
c_s^2= \frac{1}{A} \bigg [ & G_4 (K_X-2G_{3 \phi} +2 \ddot \phi G_{3X} + \dot \phi ^2 G_{3 \phi X} + \dot \phi ^2 \ddot \phi G_{3XX} +4 H \dot \phi G_{3X}) \nonumber \\
&+3G_{4 \phi}^2 - \dot \phi ^2 G_{3X} G_{4 \phi} - \frac{1}{4} \dot \phi ^4 G_{3X}^2 \bigg ], \label{cs} \\
A \equiv  G_4 \bigg [ & K_X + \dot \phi ^2 K_{XX} -2G_{3 \phi} - \dot \phi ^2 G_{3 \phi X} + 3H \dot \phi (2G_{3X}+ \dot \phi ^2 G_{3XX}) \bigg ] \nonumber \\
&+3 \left ( G_{4 \phi} - \frac{1}{2} G_{3X} \dot \phi ^2  \right )^2. \label{csd}
\end{align}
The expression (\ref{cs}) is obtained by substituting 
$\rho _\mathrm{matter} + p_\mathrm{matter}=- \rho _\phi - p_\phi -2 \dot H / \kappa ^2$ 
into Eq.~(3.12) in Ref.~\cite{DeFelice:2011bh}. 
\section{General properties of stable phantom dark energy \label{sec3}}
The no-ghost condition for scalar mode and that for tensor mode are given by $A>0$ and $G_4>0$, respectively \cite{DeFelice:2011bh}. 
Therefore, non-negativeness of the sound speeds and no-ghost conditions give the following stability conditions:  
\begin{align}
G_4 (K_X-2G_{3 \phi} +2 \ddot \phi G_{3X} + \dot \phi ^2 G_{3 \phi X} + \dot \phi ^2 \ddot \phi G_{3XX} +4 H \dot \phi G_{3X}) \nonumber \\
+3G_{4 \phi}^2 - \dot \phi ^2 G_{3X} G_{4 \phi} - \frac{1}{4} \dot \phi ^4 G_{3X}^2 \geq 0, \label{c1} \\
G_4 \bigg [ K_X + \dot \phi ^2 K_{XX} -2G_{3 \phi} - \dot \phi ^2 G_{3 \phi X} + 3H \dot \phi (2G_{3X}+ \dot \phi ^2 G_{3XX}) \bigg ] \nonumber \\
+3 \left ( G_{4 \phi} - \frac{1}{2} G_{3X} \dot \phi ^2  \right )^2 >0, \label{c2} \\
G_4>0. \label{c2.5}
\end{align}
Whereas, the conditions for realizing phantom dark energy, which are $\rho _\phi  >0$ and $w_\phi = p_\phi / \rho _\phi <-1$, are explicitly 
written as 
\begin{align}
\dot \phi ^2 K_X - K - \dot \phi ^2 G_{3 \phi} + 3 H \dot \phi ^3 G_{3X} + 3H^2 \left ( \frac{1}{\kappa ^2} -2G_4 \right ) 
-6H \dot \phi G_{4 \phi} >0, \label{c3} \\
\dot \phi ^2 K_X - \dot \phi ^2 (2 G_{3 \phi} + \ddot \phi G_{3X} -3H \dot \phi G_{3X})-2 \dot H \left ( \frac{1}{\kappa ^2} -2G_4 \right )
+2 (\ddot \phi -H \dot \phi) G_{4 \phi} +2 \dot \phi ^2 G_{4 \phi \phi} <0. \label{c4}
\end{align}
In the following, we will evaluate the conditions (\ref{c1})-(\ref{c4}) by using case analysis. 
\subsection*{The case $G_{3}(\phi , X) = G_{3}(\phi)$ and $G_4 (\phi) = 1/(2 \kappa ^2)$ }
If $G_{3}(\phi , X)$ only depends on $\phi$ and $G_4 (\phi) = 1/(2 \kappa ^2)>0$, the conditions (\ref{c1})-(\ref{c4}) are 
written as
\begin{align}
K_X -2G_{3 \phi} \geq 0, \label{c1c}\\
K_X+\dot \phi ^2 K_{XX} -2G_{3 \phi} >0, \\
\dot \phi ^2 K_X - K - \dot \phi ^2 G_{3 \phi} >0, \\
\dot \phi ^2 K_X - 2 \dot \phi ^2 G_{3 \phi} <0. \label{c4c}
\end{align}
There is a contradiction between inequalities (\ref{c1c}) and (\ref{c4c}). Therefore, 
phantom dark energy cannot be realized without instability in this case.  
This result shows that stable phantom dark energy can be only realized if 
there is a $\phi$ dependence in $G_4$ or a $X$ dependence in $G_3$. 
The reason why we do not consider constant $G_4 $ which is different from $1/(2 \kappa ^2)$ is 
in order to consistent with the solar system tests and the laboratory experiments of gravitation. In fact, $O(10^{-5})$ difference in the value of $G_4$ 
is only allowed by the experiments \cite{Mohr:2015ccw}, however, 
such a small difference does not affect the conditions (\ref{c1c})-(\ref{c4c}). 
\subsection*{The case $G_3$ has a $X$ dependence}
If $G_{3}$ has a $X$ dependence and $G_4 (\phi) = 1/(2 \kappa ^2)$, the conditions (\ref{c1})-(\ref{c4}) are 
written as
\begin{align}
K_X -2G_{3 \phi} + 2 \ddot \phi G_{3X} + \dot \phi ^2 G_{3 \phi X}+ \dot \phi ^2 \ddot \phi G_{3XX} 
+ 4H \dot \phi G_{3X} - \frac{\kappa ^2}{2} \dot \phi ^4 G_{3X}^2 \geq 0, \label{c1x}\\
K_X+\dot \phi ^2 K_{XX} -2G_{3 \phi}- \dot \phi ^2 G_{3 \phi X} 
+3H \dot \phi (2G_{3X}+ \dot \phi ^2 G_{3XX})+\frac{3 \kappa ^2}{2} \dot \phi ^4 G_{3X}^2>0, \label{c2x}\\
\dot \phi ^2 K_X - K - \dot \phi ^2 G_{3 \phi}+3H \dot \phi ^3 G_{3X} >0, \label{c3x}\\
\dot \phi ^2 K_X - \dot \phi ^2 (2G_{3 \phi} + \ddot \phi G_{3X} - 3H \dot \phi G_{3X}) <0. \label{c4x}
\end{align}
Both inequalities (\ref{c1x}) and (\ref{c4x}) can be satisfied only if 
\begin{equation}
\left ( 3 \ddot \phi + H \dot \phi - \frac{\kappa ^2}{2} \dot \phi ^4 G_{3X} \right ) G_{3X} + \dot \phi ^2 (G_{3 \phi X}+ \ddot \phi G_{3XX}) >0. 
\end{equation}
Inequalities (\ref{c1x})-(\ref{c4x}) are so complicated that it is difficult to find appropriate function forms of 
$K(\phi ,  X)$ and $G_3(\phi , X)$ which satisfy all of the conditions, 
however, we can find the large/small relations between the functions in the following manner. 
In the case of potential driven slow-roll accelerated expansion 
i.e. the case $K(\phi , X)=X-V( \phi )$, $X \ll V(\phi) \sim 3H^2/ \kappa ^2$, and $| \ddot \phi | \ll | H \dot \phi|$, inequality (\ref{c3x}) is automatically satisfied 
because $V(\phi)$ is much larger than the other terms. 
While, inequalities (\ref{c1x}) and (\ref{c4x}) imply $H \dot \phi G_{3X} $ or the other terms proportional to $G_{3X}$ or $G_{3 \phi X}$ should be 
same order as $K_X- 2G_{3 \phi}=1- 2G_{3 \phi}$, 
because the terms $K_X - 2G_{3 \phi}$ commonly exist  in inequalities (\ref{c1x}) and (\ref{c4x}) but 
the signs of inequalities are different. In particular, if $G_{3 \phi X}$ and $G_{3 XX}$ are negligible and $K_X- 2G_{3 \phi}$ is negative, 
inequalities (\ref{c1x}) and (\ref{c4x}) can be satisfied by positive $H \dot \phi G_{3X} $ which is larger than $-(K_X- 2G_{3 \phi})/4$ 
and less than $-(K_X- 2G_{3 \phi})/3$. 

\subsection*{The case $G_{3}(\phi , X) = 0$ and $G_4 (\phi) $ has a $\phi$ dependence }
If we only take $K(\phi , X)$ and $G_4 (\phi)$ into account, then the conditions (\ref{c1})-(\ref{c4}) are 
given as
\begin{align}
G_4 (K_X G_4 + 3 G_{4 \phi}^2) \geq 0, \label{c1f}\\
G_4 [ G_4 (K_X + \dot \phi ^2 K_{XX}) + 3 G_{4 \phi} ^2 ] >0, \label{c2f}\\
G_4>0, \label{c2.5f} \\
3 \left ( \frac{1}{\kappa ^2}-2 G_4  \right ) H^2 + \dot \phi ^2 K_X - K -6H \dot \phi G_{4 \phi} >0, \label{c3f}\\
-2 \left ( \frac{1}{\kappa ^2}-2 G_4  \right) \dot H + \dot \phi ^2 K_X +2 (\ddot \phi - H \dot \phi ) G_{4 \phi} + 2 \dot \phi ^2 G_{4 \phi \phi} <0. \label{c4f}
\end{align}
In this case, to realize stable dark energy is much easier than that in the case that $G_{3}$ has a $X$ dependence and $G_4 (\phi) = 1/(2 \kappa ^2)$, because 
inequalities (\ref{c1f})-(\ref{c2.5f}) are always established as long as $G_4>0$, $K_X>0$, and $K_{XX} \geq 0$. 
In particular, in the case of canonical kinetic term, which is $K(\phi ,  X) = X- V(\phi)$, inequalities (\ref{c1f})-(\ref{c2.5f}) are completed if $G_4>0$, 
moreover, sound speed of scalar mode $c_s$ always satisfies $c_s^2 =1$. 
Then, we can adjust two arbitrary functions $V(\phi)$ and $G_4 (\phi)>0$ for making the conditions (\ref{c3f}) and (\ref{c4f}) true. 
\section{Examples \label{sec4}}
\subsection*{The case $G_{3}$ has a $X$ dependence and $G_4 (\phi) = 1/(2 \kappa ^2)$ }
\begin{figure}
\begin{minipage}[t]{0.5\columnwidth}
\begin{center}
\includegraphics[clip, width=0.97\columnwidth]{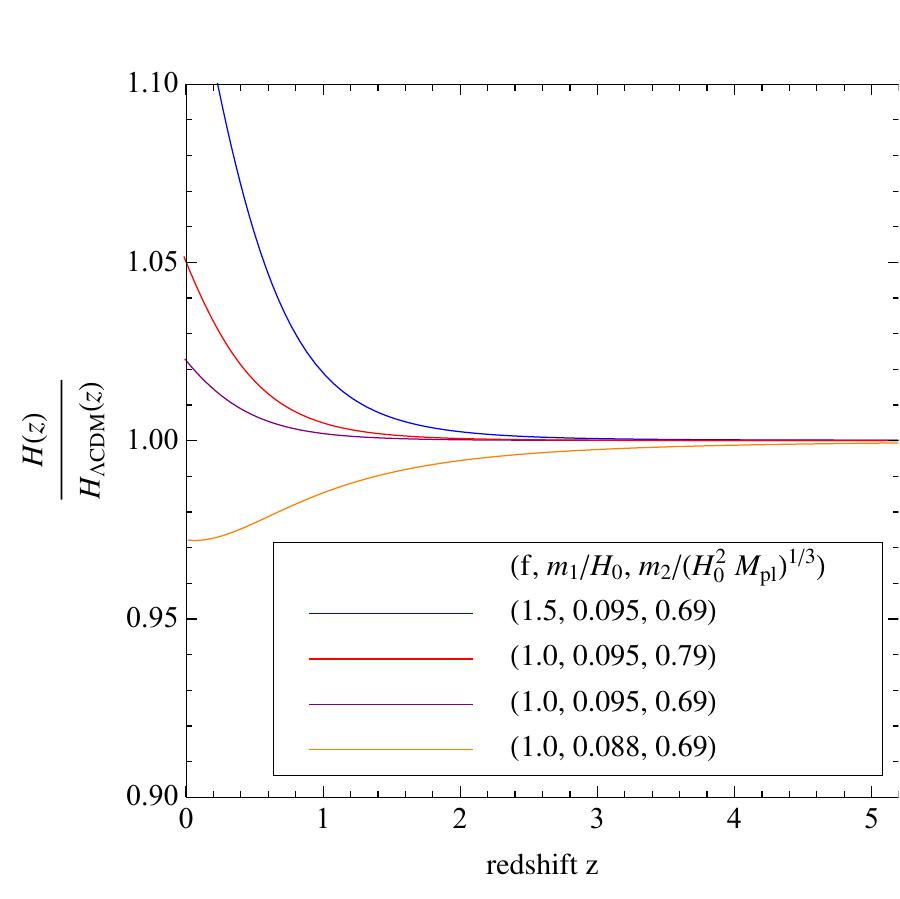}
\end{center}
\end{minipage}%
\begin{minipage}[t]{0.5\columnwidth}
\begin{center}
\includegraphics[clip, width=0.97\columnwidth]{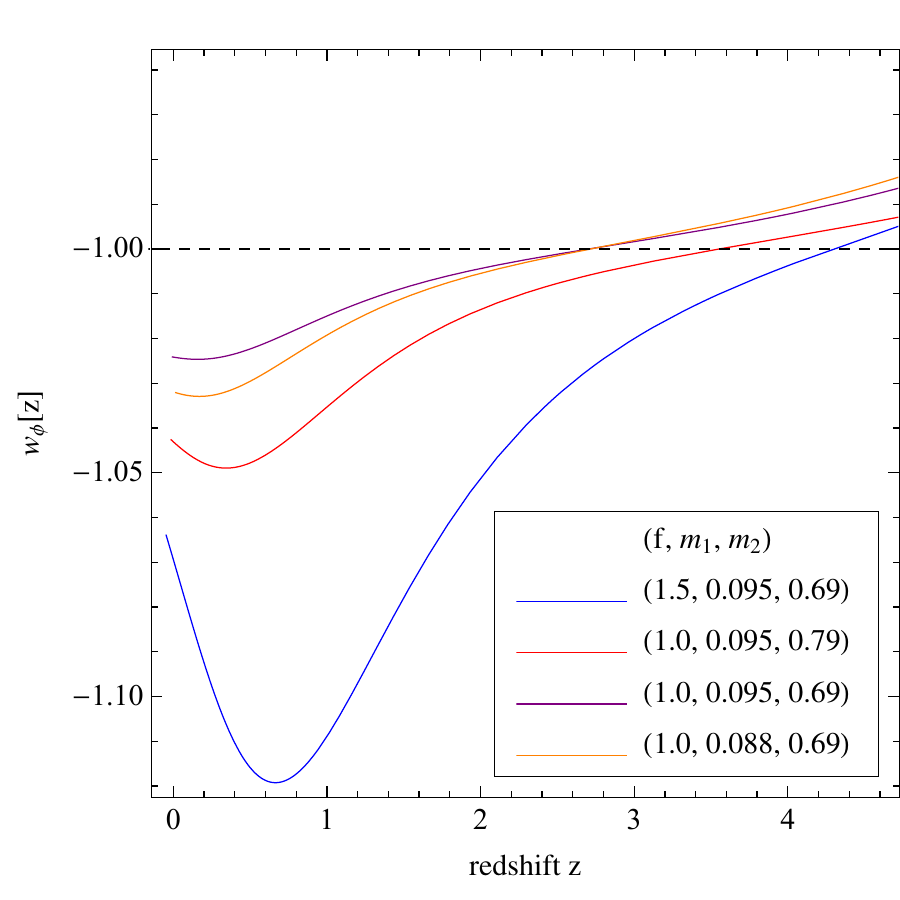}
\end{center}
\end{minipage}
\caption{Redshift dependence of expansion rate compared to the $\Lambda$CDM model (left) and that of the effective 
equation of state parameter of the scalar field $w_\phi = p_\phi / \rho _\phi$ (right)
in the case of $K(\phi ,  X) = X- m_1^2 \phi ^2$ and $G_3(\phi ,X)=f \phi + X/m_2^3$.  
The cases; higher value in $f$, higher value in $m_2$, and lower value in $m_1$, are 
expressed as Blue curve, Red curve, and Orange curve, respectively. 
The initial conditions for $\phi (z)$ and $\dot \phi (z)$ are assigned as $\phi (10)=3M_{pl}$ and $\dot \phi (10)=0.04 M_{pl}H_0$, 
where $H_0$ means that the Hubble constant in the $\Lambda$CDM model: $H_0 \simeq 68$ (km/s)/Mpc. 
And $\Omega _{\mathrm{matter},0}=0.31$ is assumed to plot the figures. 
}
\label{g3h}
\end{figure}
As shown in the previous section, slow-rolling scalar field has a possibility to be a stable phantom dark energy. 
In the case of 
\begin{equation}
K(\phi ,  X) = X- m_1^2 \phi ^2 \quad and \quad G_3(\phi ,X)=f \phi +\frac{X}{m_2^3}, 
\end{equation} 
where $m_1$, $m_2$, and $f$ are positive constants, then $G_{3 \phi}, G_{3 X} >0$ and $G_{3 \phi X}= G_{3 XX} =0$ are satisfied, 
so the large/small relation written in the previous section can be realized by choosing appropriate values of $m_2$ and $f$. 
\begin{figure}
\begin{minipage}[t]{0.5\columnwidth}
\begin{center}
\includegraphics[clip, width=0.97\columnwidth]{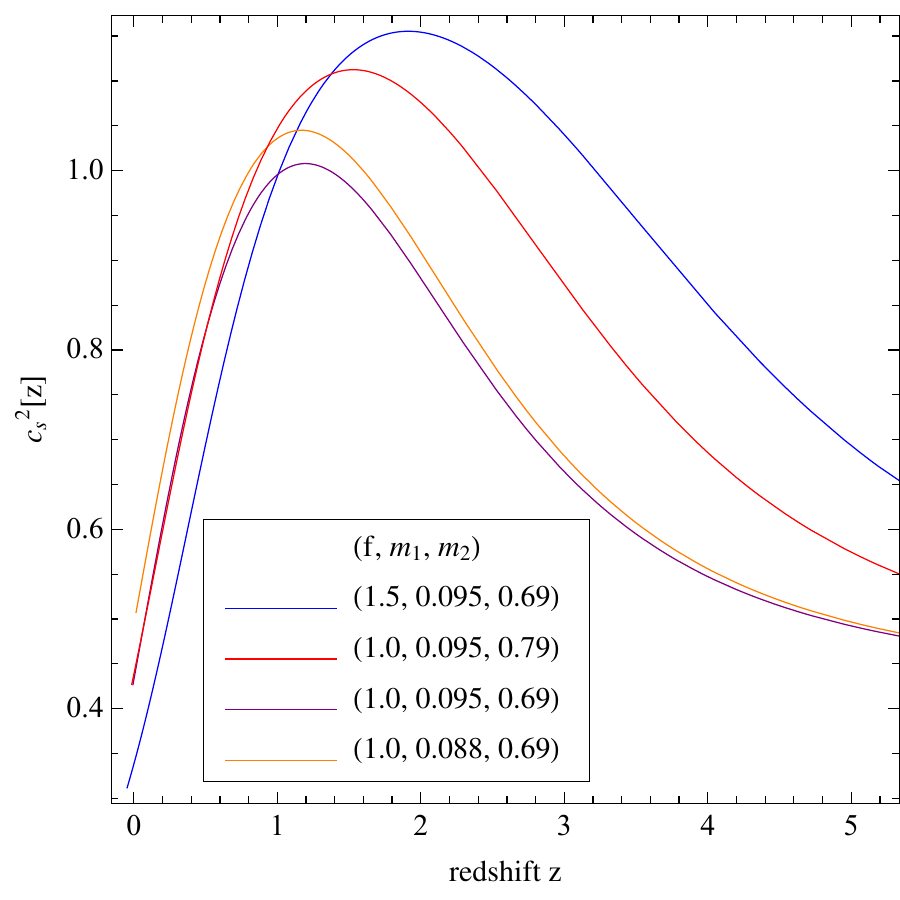}
\end{center}
\end{minipage}%
\begin{minipage}[t]{0.5\columnwidth}
\begin{center}
\includegraphics[clip, width=0.97\columnwidth]{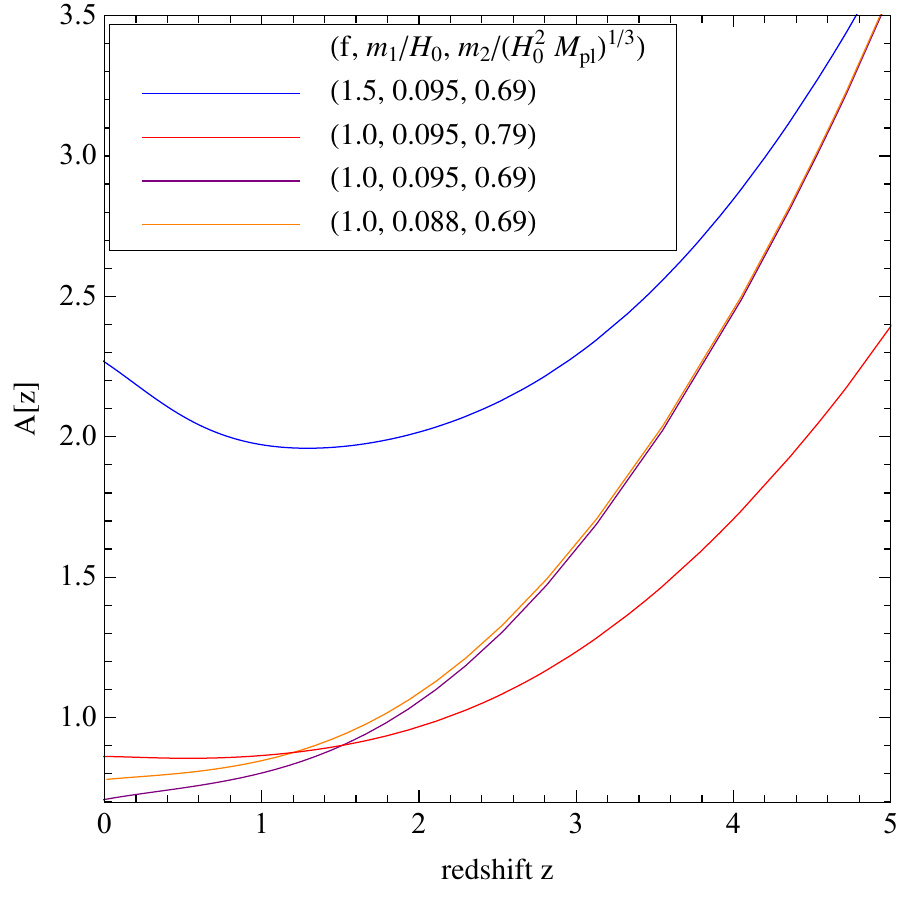}
\end{center}
\end{minipage}
\caption{Redshift dependence of $c_s^2$ and $A$ in the case of $K(\phi ,  X) = X- m_1^2 \phi ^2$ and $G_3(\phi ,X)=f \phi + X/m_2^3$.  
Initial conditions are same as those in Fig.~\ref{g3h}}. 
\label{g3cs}
\end{figure}
In Fig.~\ref{g3h}, redshift dependence of the Hubble rate parameter and that of the effective equation of state parameter of the scalar field, 
which is defined as $w_\phi = p_\phi / \rho _\phi$, are depicted. 
As seen in the right figure, equation of state of the scalar field crosses over the phantom divide, which is the boundary $w_\phi =-1$, around $z=3$. 
By the effect of phantom crossing, the Hubble rate becomes greater than that of the $\Lambda$CDM model except for Orange curve. 
In the case of Orange curve, dark energy density is less than that of the other curves, therefore, the Hubble rate cannot be greater than that of the $\Lambda$CDM model 
in small redshift region. 
In Fig.~\ref{g3cs}, evolution of $c_s^2$ and that of $A$ defined in Eq.~(\ref{csd}) are depicted. 
Both $c_s^2$ and $A$ are always positive, therefore, there are not instability at least in the region $0<z<5$. 
However, we should be careful that $w _\phi < -1$ means the break down of the null energy condition because the effective equation of state parameter of the scalar field 
$w_ \phi$ is not only "effective" but also "exact" in the case $G_4 (\phi) = 1/(2 \kappa ^2)$ and $G_5 (\phi , X)=0$. 

\subsection*{The case $G_{3}(\phi , X) = 0$ and $G_4 (\phi) $ has a $\phi$ dependence }
\begin{figure}
\begin{minipage}[t]{0.5\columnwidth}
\begin{center}
\includegraphics[clip, width=0.97\columnwidth]{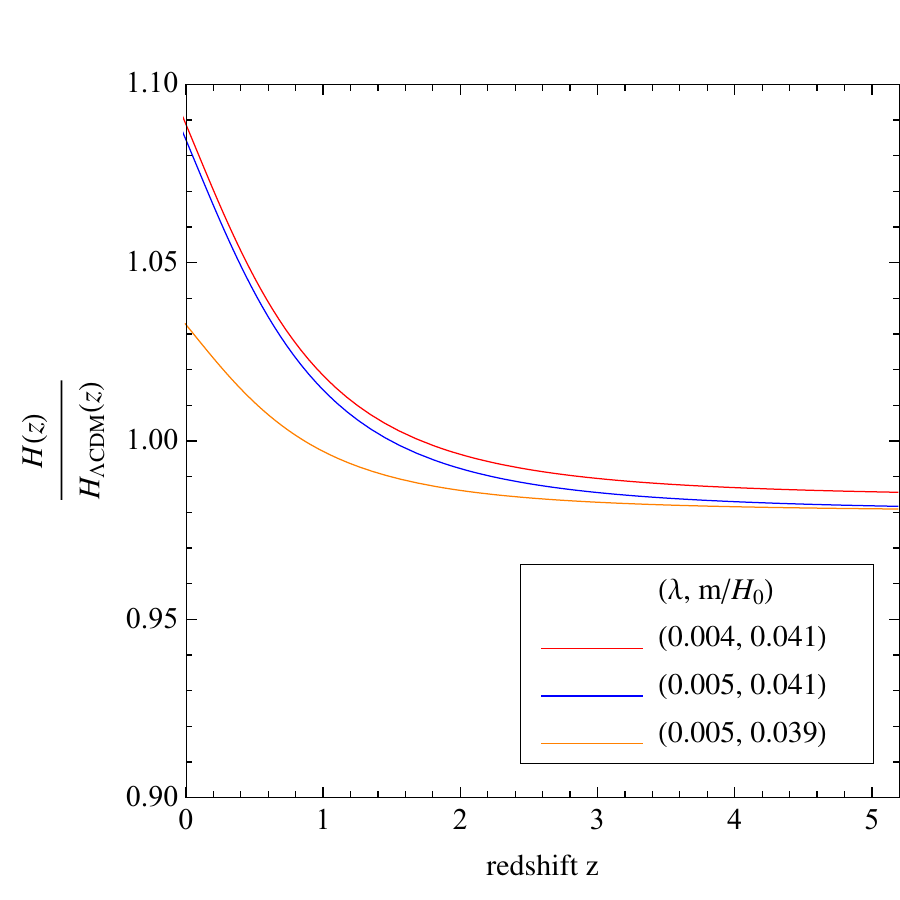}
\end{center}
\end{minipage}%
\begin{minipage}[t]{0.5\columnwidth}
\begin{center}
\includegraphics[clip, width=0.97\columnwidth]{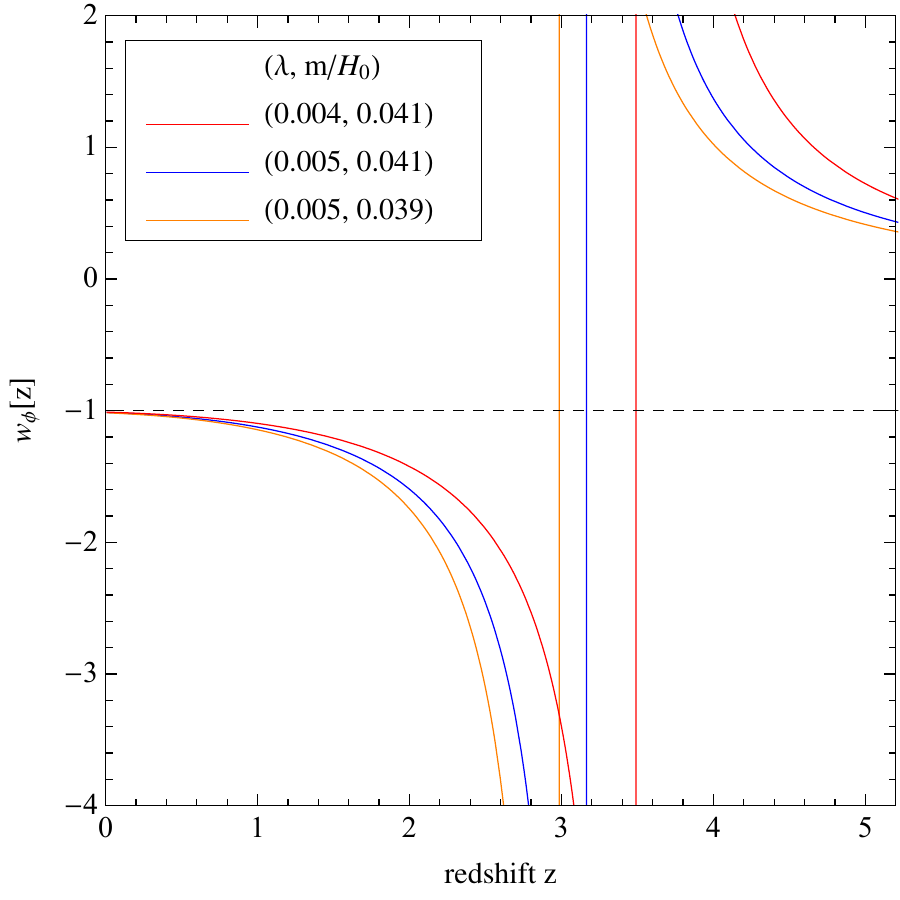}
\end{center}
\end{minipage}
\caption{Redshift dependence of the Hubble rate function (left) and that of the effective equation of state parameter of the scalar field (right) in case of 
$K(\phi ,  X) = X- m^2 \phi ^2$ and $G_4(\phi )= \frac{1}{2 \kappa ^2} \mathrm{e}^{ \lambda \phi /M_{pl}}$. 
Red curve and Orange curve express a lower value in $\lambda$ and a lower value in $m$, respectively. 
The initial conditions for $\phi (z)$ and $\dot \phi (z)$ are assigned as $\phi (10)=8M_{pl}$ and $\dot \phi (10)=0.04 M_{pl}H_0$.}
\label{g4h1}
\end{figure}

Let us first consider the slow-roll accelerated expansion of the Universe. If $K(\phi ,  X) = X- V(\phi)$, $X \ll V(\phi) \sim 3H^2/ \kappa ^2$, and $| \ddot \phi | \ll | H \dot \phi|$, 
then the inequalities (\ref{c3f}) and (\ref{c4f}) can be satisfied for $G_4 \simeq 1/(2 \kappa ^2)$, $G_{4 \phi} >0$, and $\dot \phi >0$, 
because the term $V(\phi)$ is the dominant component in the left-hand-side of (\ref{c3f}), and $-2H \dot \phi G_{4 \phi}$ can be dominant in the 
left-hand-side of (\ref{c4f}) if $|G_{4 \phi \phi}|  \lesssim |G_{4 \phi}/M_{pl}|$. Such a situation is realized by assuming 
\begin{equation}
K(\phi ,  X) = X- m^2 \phi ^2 \quad and \quad G_4(\phi )= \frac{1}{2 \kappa ^2} \mathrm{e}^{ \lambda  \frac{\phi }{ M_{pl}}}, \label{g4n1}
\end{equation}
with positive $\lambda$. In Fig.~\ref{g4h1}, the evolution of the Hubble rate function and the effective equation of state parameter of the 
scalar field in case of Eq.~(\ref{g4n1}) are expressed. 
The phantom crossing is realized around $z=3$ as seen in the right figure. 
This phantom crossing is caused not by the change in $p_\phi$ but by the inversion of the sign of $\rho _\phi$. 
Here, negative $\rho _\phi$ does not mean the existence of negative energy, because $\rho _\phi$ not only contains usual energy density of the scalar field 
but also contains the deviations from the Einstein gravity. This type of phantom crossing is also seen in the other modified gravity models \cite{Sahni:2014ooa,Sahni:2002dx}. 
One may think that the discontinuity in the equation of state parameter have a great effect on the observational quantities. 
In fact, the effect is small because there are not steep changes in the quantities $\rho _ \phi$ and $p _ \phi$, moreover, dark energy density is much smaller than 
that of dark matter around $z=3$. 
Therefore, the effects on $H$ and $\dot H$ from the discontinuity in the equation of state are negligibly small. 
In the left figure, we can see that the Hubble rate in this case is a bit smaller than that in the $\Lambda$CDM model in large $z$. This is caused from 
the modification of gravity, namely, deviation from $G_4 (\phi)  \equiv 1/(2 \kappa ^2)$ effectively play a role of negative energy. However, in low redshift region, the Hubble rate 
which is larger than $H_{\Lambda CDM}$ is realized by the effect of mass term $m^2 \phi ^2$. 

\begin{figure}
\begin{minipage}[t]{0.5\columnwidth}
\begin{center}
\includegraphics[clip, width=0.97\columnwidth]{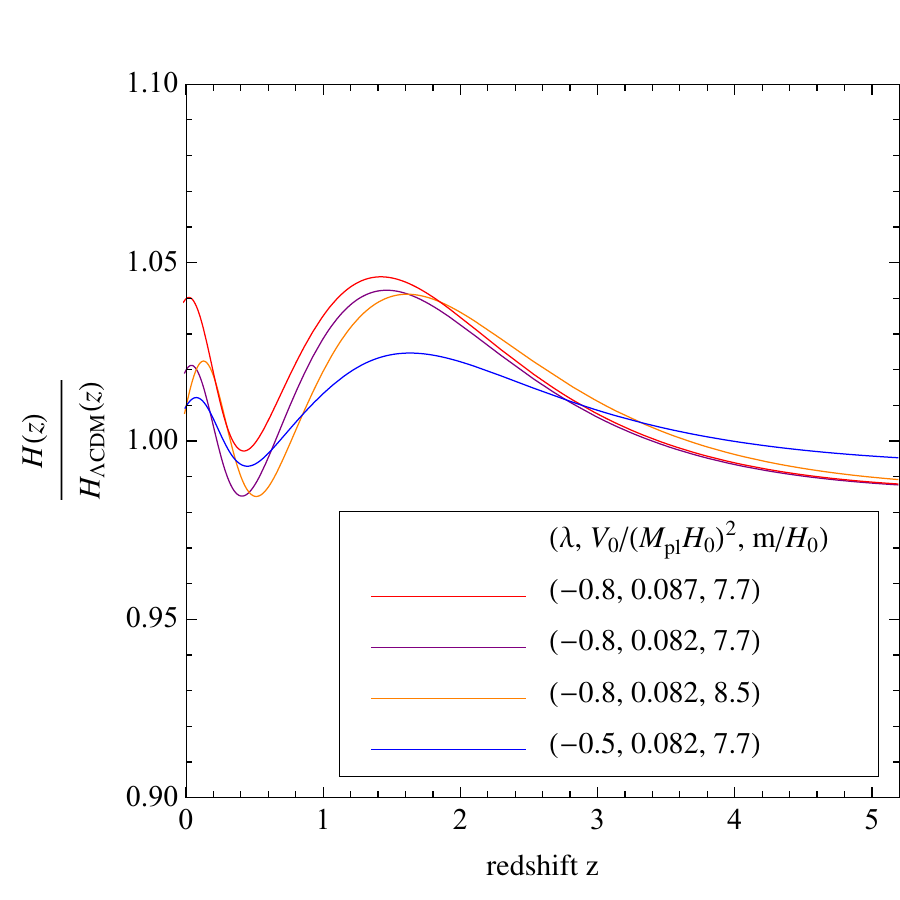}
\end{center}
\end{minipage}%
\begin{minipage}[t]{0.5\columnwidth}
\begin{center}
\includegraphics[clip, width=0.97\columnwidth]{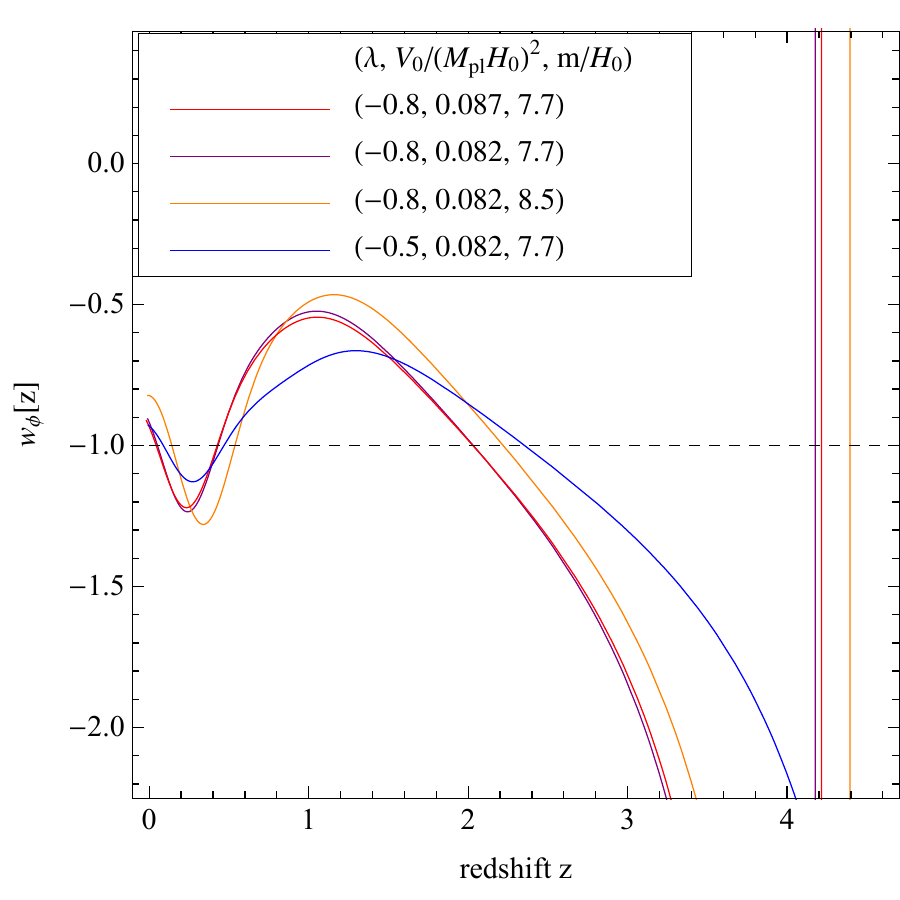}
\end{center}
\end{minipage}
\caption{Redshift dependence of the Hubble rate function (left) and that of the effective equation of state parameter of the scalar field (right) in case of 
$K(\phi ,  X) = X- (V_0+m^2 \phi ^2)$ and $G_4(\phi )= \frac{1}{2 \kappa ^2} \mathrm{e}^{ \lambda \phi /M_{pl}}$. 
Red curve, Orange curve, and Blue curve express a higher value in $V_0$, a higher value in $m$, 
and a higher value in $\lambda$, respectively. 
The initial conditions for $\phi (z)$ and $\dot \phi (z)$ are assigned as $\phi (10)=-0.03M_{pl}$ and $\dot \phi (10)=0.04 M_{pl}H_0$.}
\label{g4h2}
\end{figure}
\begin{figure}
\begin{center}
\includegraphics[clip, width=0.70\columnwidth]{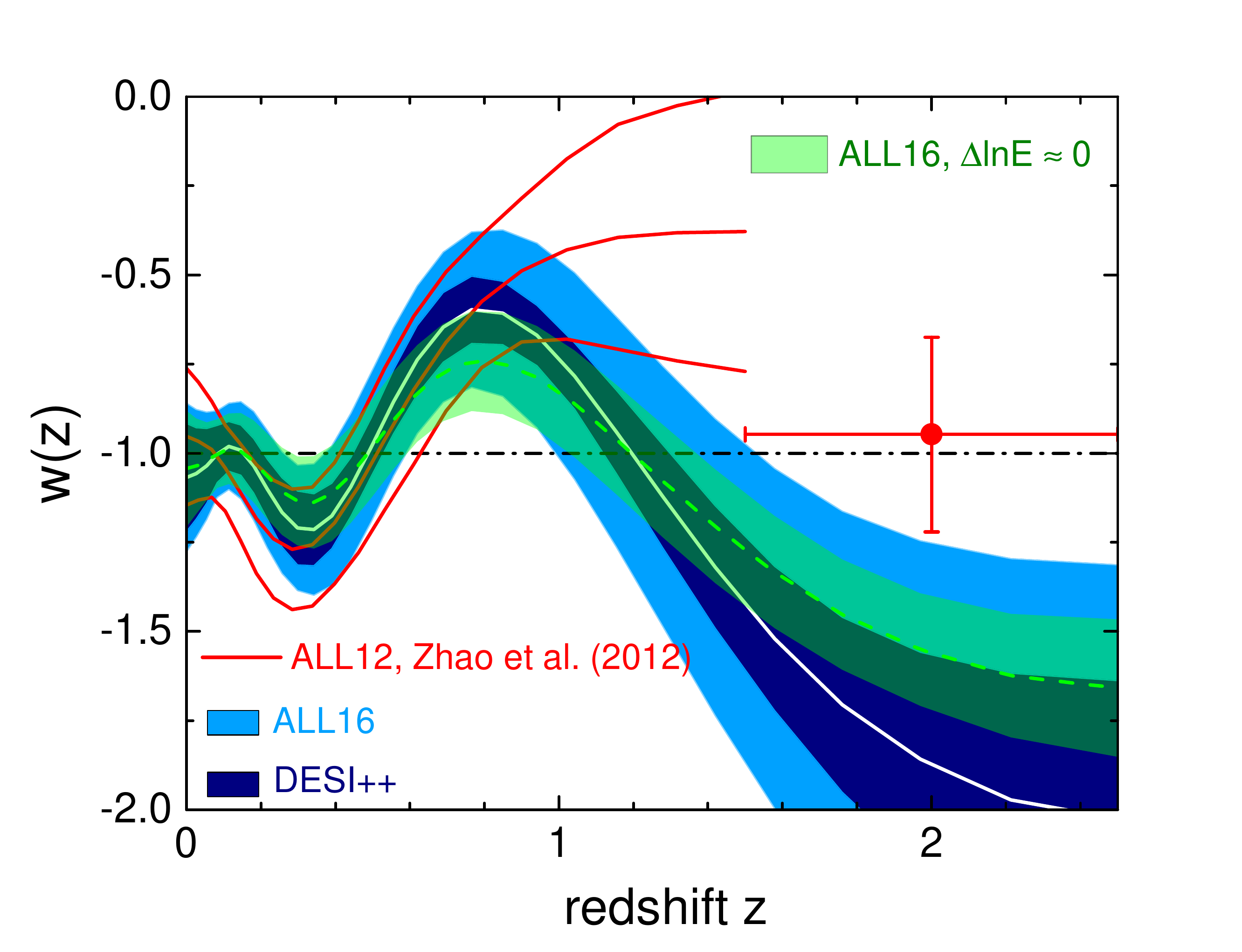}
\end{center}
\caption{"The reconstructed evolution history of the dark energy equation of state compared with the 2012 
result and the forecasted uncertainty from future data" from Ref.~\cite{Zhao:2017cud}.}
\label{rew}
\end{figure}

If we generalize the potential given in Eq.~(\ref{g4n1}), we can describe more complicated behavior of $w_\phi$. 
In the case of 
\begin{equation}
K(\phi ,  X) = X- (V_0+m^2 \phi ^2) \quad and \quad G_4(\phi )= \frac{1}{2 \kappa ^2} \mathrm{e}^{\lambda  \frac{\phi }{ M_{pl}}}, \label{g4n2}
\end{equation}
where $V_0$ is a positive constant of mass dimension four, 
$V_0$ cause the accelerated expansion of the Universe same as the case of the $\Lambda$CDM model. 
However, the effects from $m^2 \phi ^2$ and $ \mathrm{e}^{\lambda  \phi / M_{pl}}$ can dramatically change the behavior 
of $H(z)$ and $w_\phi (z)$ compared to the $\Lambda$CDM model. 
In Fig.~\ref{g4h2}, the evolution history of $H(z)$ and $w_\phi (z)$ are depicted. 
In the same matter as the case (\ref{g4n1}), a phantom crossing caused by the inversion of the sign of $\rho _\phi$ occur around $z=4$. 
However, several phantom crossing happen after it in this case. This is because the field $\phi$ oscillates around the stationary point. 
The oscillation of $\phi$ is mainly caused by the mass term $m^2 \phi ^2$ and the phantom crossing is accompanied by the oscillation 
because of the factor $ \mathrm{e}^{\lambda  \phi / M_{pl}}$. 
In the figure, the difference between Purple curve and Red curve is only the value of $V_0$. 
Therefore, the value of $H(z)$ is simply increased in Red curve compared to Purple one, while, 
the behavior of $w_\phi (z)$ in Red curve is almost same as that in Purple curve. 
The reason why there is only a little difference in $w_\phi (z)$ is that the value of $w_\phi (z)$ is not influenced by the 
total amount of dark energy but influenced by the ratio between $V_0$ and the other terms. 
Interestingly, the reconstructed equation of state parameter of dark energy from the observations 
given by G.~B.~Zhao {\it et al.} \cite{Zhao:2017cud} 
have a same behavior as Fig.~\ref{g4h2} (see Fig.~\ref{rew}). There are 
phantom crossings in the reconstructed $w(z)$, therefore, the authors of the paper mention that such a 
behavior could be explained by Quintom scenario \cite{Feng:2004ad,Cai:2009zp} or the interaction between dark energy and dark matter \cite{Das:2005yj}. 
While, Fig.~\ref{g4h2} explicitly show the behavior of the reconstructed $w(z)$ is realized without instability if we consider 
Eq.~(\ref{g4n2}). 

\section{Conclusion \label{sec5}}
We have investigated the conditions in order to realize phantom dark energy without instability in Horndeski's theory. 
First, we have assumed $G_4(\phi , X)=G_4 (\phi)$ and $G_5(\phi , X) =0$ by following to the observational results of 
gravitational wave GW170817 and its electro-magnetic counterparts, then, 
there are three arbitrary functions $K(\phi ,X)$, $G_3(\phi , X)$, and $G_4(\phi)$ in Horndeski's theory. 
Under this condition, we have derived the following results.   
$X$ dependence in $G_3$ function or $\phi$ dependence in $G_4$ function 
must exist for realizing stable phantom dark energy, because there is a contradiction among the conditions 
$c_s^2 \geq 0$, $A>0$, and $w_\phi <-1$ if $G_3 (\phi , X)=G_3 (\phi)$ and $G_4 (\phi) = const.$. 
In both cases $G_3 (\phi , X) \neq G_3 (\phi)$ and $G_4 (\phi) \neq const.$, 
slow-roll accelerated expansion with mass term of the scalar field $m^2 \phi ^2$ can yield 
dark energy which cross the phantom divide. 
Moreover, it has been also shown that a behavior of $w_\phi (z)$ which is similar to 
observationally reconstructed evolution history of the dark energy equation of state \cite{Zhao:2017cud} 
can be realized in the model (\ref{g4n2}).

\section*{Acknowledgments}
The author would like to thank Frederico Arroja for discussions. The author's researches 
are supported by the Leung Center for Cosmology and Particle Astrophysics (LeCosPA) 
of the National Taiwan University (NTU). This work is partly supported by 
the Russian Government Program of Competitive Growth of Kazan Federal University.

\end{document}